Evidence for two electronic components in $Na_xCoO_2$ ( $x$ = 0.7-0.75)

M. Brühwiler, B. Batlogg, S. M. Kazakov and J. Karpinski

*Solid State Physics Laboratory, ETH Zürich, CH-8093 Zürich, Switzerland*

Thermodynamic and transport measurements on $Na_xCoO_2$ ( $x \sim$ 0.7-0.75) over a wide temperature range reveal a strongly enhanced low energy excitation spectrum. The zero-field specific heat and resistance at low temperature are proportional to $T^n$ with $n<1$, in sharp contrast to behavior found in ordinary Landau Fermi-liquid metals. For temperatures below ~5 K this unusual excitation seen in the specific heat is partially suppressed in a magnetic field, following a $T/B$ scaling. The specific heat reduction at low $T$ is partially compensated by an increase at higher temperature. A quantitative comparison with the magnetoresistance and the thermopower indicates that the low-energy electronic state requires a description beyond the single band model.

The crystal structure of $Na_xCoO_2$ is characterized by $CoO_2$ layers separated by the Na atoms, with the Co atoms forming a triangular lattice due to the edge-sharing of the Co-O octahedra in the layers [1-3]. The large thermoelectric power [4-8] and the anomalous high-temperature Hall effect [9] are signs of an unusual electronic state. The recently discovered superconductivity in water-intercalated $Na_{0.35}CoO_2 \cdot yH_2O$ [10], studied by several groups [11-21], has drawn further attention to this layered transition-metal oxide. The thermo-power at low temperatures can be suppressed by a magnetic field, indicating that spin entropy significantly contributes to the thermopower [22]. This compound is of interest also from a conceptual point of view, as it realizes a triangular lattice with partially filled strongly correlated electronic states [23-29]. A rich electronic phase diagram and an unusual linear temperature dependence of the high-temperature Hall effect has been predicted [e.g. in 24], and such a Hall effect has been measured above ~200 K [9]. Electronic band structure calculations for $Na_{0.5}CoO_2$ give two sets of bands crossing the Fermi level. One main hole-like band around the center of the Brillouin zone, and six small hole-like pockets further out towards the zone boundary [30]. Angle-resolved photoemission spectroscopy supports the existence of a large hole-like band [31].

To study the electronic state of $Na_xCoO_2$, and particularly the development of the quantum ground state, we have measured the specific heat and electrical transport of a number of samples in single and polycrystalline form from 0.5 K to 320 K, in magnetic fields up to 14 Tesla. At low temperature we measure a strongly enhanced and rapidly changing resistance $R(T)$ and specific heat $C(T)$. The zero-field $R(T)$ and $C(T)$ follow a highly unusual temperature dependence, distinctly different from that of an ordinary metal. Magnetic fields have a significant influence mainly on the specific heat. Comparing the influence of a magnetic field on the resistance, the specific heat and on the thermopower [9], leads to the conclusion that the low temperature properties are determined by the presence of two distinct electronic components. One of them involves a low energy scale and is strongly affected by a magnetic field, while the other is of a conventional electronic quasiparticle nature, albeit renormalized by correlations, and is not significantly affected by a magnetic field. Their relative contribution to the measured low temperature properties is significantly different.

The $Na_xCoO_2$ ($x$=0.7, 0.75) powder samples were prepared employing the conventional synthesis method involving either slow heating or the "rapid heat-up" technique [6]. A stoichiometric mixture of $Co_3O_4$ (Aldrich, 99.995%) and $Na_2CO_3$ (Aldrich, 99.995%) was pressed in pellets and annealed overnight at 750-850 °C in air. The samples were confirmed by X-ray powder diffraction to be a single phase of hexagonal $\gamma$-$Na_xCoO_2$ [1,2]. Single crystals were grown from NaCl flux according to the procedure described in [4]. Specific heat and resistance were measured in a physical properties measurement apparatus (Quantum Design, PPMS). For an additional cross-check an identical sample was measured on different sample holders with different thermometers, and excellent agreement was found between the results.

The resistance $R(T)$ of several samples with x=0.7-0.75 is shown in Figure 1 normalized to the value at 200 K. Anomalies observed at higher temperatures will be discussed separately. Upon cooling, $R(T)$ decreases as in typical metals down to about 100 K, and then develops a broad hump indicative of enhanced scattering. A broad hump is also seen in single crystals when the current flows parallel to the $CoO_2$ layers (top solid line in Fig. 1), but the resistance tends to saturate at lowest $T$. The resistance for the polycrystals reflect the resistance in the $CoO_2$ layers, as the resistance perpendicular to the layers is larger by orders of magnitudes [4,7]. (A $T^{1.5}$ temperature dependence for the in-plane resistance has been reported for a $Na_{0.58}CoO_2$ crystal [33]).

Particularly remarkable is the continuously increasing slope $dR/dT$ towards low temperatures observed in all polycrystalline samples (Fig. 1 and Fig. 2). In all polycrystalline samples $R(T)$ continues to decrease to at least 2 K, with a drop of up to ~ 20 – 50 % from 5 K to 2 K. The measured value at 2 K is as low as ~ 4 % of $R(200K)$, and a reasonable extrapolation to zero temperature yields a residual-resistance-ratio $R_{300}/R_0$ up to 50 - 100. The zero-field temperature dependence of $R(T)$ can be parametrized by a power law $R(T) = R_0 + AT^n$ over an extended temperature range up to ~ 30 – 40 K. The increase of $R(T)$ is sub-linear in T with $n \sim$ 0.65 – 0.7 for all samples. This peculiar $T$-dependence reflects a highly unusual and strong scattering of the charge excitations with a rapidly decreasing scattering towards zero temperature.

Measurements of the specific heat provide further evidence of the unusual excitation spectrum. (Lower panel in Fig. 2). The over-all values reflect significant mass enhancement compared to ordinary metals, due to correlation among the d-electrons near the Fermi level [30]. In the range of overlap, the data of all five samples measured in this study are similar to previously reported results [32]. Instead of following a combination of a phononic ($C \propto T^3$) and usual electronic contribution ($C \propto T$), the zero-field $C(T)$ approaches zero Kelvin as $C \propto T^n$ with $n < 1$ (up to $T$ ~ 7 K, $n = 0.8 – 0.85$, broken line in Fig. 2). This sub-linear $T$ dependence indicates an increasing density-of-states (DOS) at low energies, and leads to an upturn in $C/T$ below 5-8 K (Fig. 3).

To summarize the key findings, $R(T)$ and $C(T)$ are compared in Fig. 2, where the $T$ scales have been chosen to account for the different energy windows probed when measuring the two different physical properties. Both quantities follow a similarly unusual temperature dependence and reflect the peculiar nature of the electronic state in this metal.

Valuable insight into the electronic energy scales can be gained by comparing the specific heat results with band structures from LDA calculations and with a simple model band structure. If the specific heat at low temperature is interpreted in terms of an electronic contribution "$\gamma T$", ignoring for a moment the upturn in $C/T$, and assuming a flat DOS within a few meV, the value of "$\gamma$" for all samples is in the range of ~25-30 mJ/mol-K$^2$. The LDA band structure value of ~ 11 mJ/mol-K$^2$ was calculated for Na$_{0.5}$CoO$_2$, and ~7 mJ/mol-K$^2$ can be estimated from moving E$_F$ within a rigid band [30]. Thus the measurements give an enhancement of the DOS by a factor of 3-4. The measured "$\gamma$" can also be compared with the DOS of a model tight binding band structure for a triangular lattice [24,28], parametrized by an effective hopping integral $t_{eff}$ whose sign is currently under discussion. If $t_{eff} > 0$ is assumed as in [24] we estimate for the present band filling a DOS of ~ 0.5 states/$t_{eff}$ and calculate with the measured "$\gamma$" a $t_{eff}$ ~39 - 45 meV, and a Fermi energy of ~ 48 – 58 meV. If $t_{eff} < 0$ is assumed, as e.g. in Ref. [19] following band structure calculations, the DOS is ~ 0.11 states/$t_{eff}$ for x~0.7 and gives a $t_{eff}$ ~ 9-11 meV. The overall bandwidth of $9 t_{eff}$ (~ 80-100 meV) would be ~ 14 to 17 times smaller than the LDA band width of ~1.4 eV. The corresponding Fermi energy (for holes) is ~ 21-27 meV. This value for $E_F$ is in remarkable agreement with the Hall-effect studies, where $R_H$ is found to increase linearly above ~ 200 K, representing a scale of the degeneracy temperature [9]. We note that the measured $\gamma$ values and the estimated $t_{eff}$ are about twice as large in the non-superconducting samples with a Na content $x$ ~0.7 (also in Ref. [32]) than in the superconducting ones with $x$ ~0.35. [17,19,20]. The difference may be taken as to indicate that the renormalization of the bandwidth depends on the band filling, and/or that the real band structure is not captured in all details by the single band model. It would be interesting to compare the results with the calculations for the Kagomé lattice hidden in the triangular lattice of $CoO_2$ [29].

To further study the electronic state, we have measured the specific heat in magnetic fields up to 14 T. The changes of $C(T,B)$ with field are pronounced and are neither monotonic in field or temperature. In Fig. 3 $C(T,B)/T$ is shown for magnetic fields up to 14 T, in steps of 2 T. A close inspection of the data at lowest temperatures and highest fields reveals a contribution to $C(T,B)$ consistent with a Schottky anomaly, i.e. proportional to $(B/T)^2$ with a prefactor of 0.011 mJ K / mol-T². For the following discussion and the plots in Figs. 3 and 4, this part has been subtracted from our data. Two regions can be distinguished. For $T$ of a few Kelvin, the specific heat is suppressed in a magnetic field. The suppression of low energy fluctuations is partly compensated by an increase of $C/T$ at higher $T$, although it appears that the entropy is not balanced in the temperature range covered in the measurements. The reduction of the excitations at low energy is the larger the higher the magnetic field, and it follows rather well a $T/B$ scaling (inset Fig. 3). The increase of $C/T$ at higher $T$, however, is not found to scale with $T/B$. The analogy to the $B/T$ scaling observed for the suppression of the thermoelectric power $\Delta Q/Q$ [22] is apparent, as one would expect for thermodynamic reasons. However, notable differences are discussed in the following.

A quantitative comparison of the relative changes of $Q(T,B)$, $C(T,B)$ and $R(T,B)$ is revealing (Fig. 4). In the maximum field of 14 T the largest suppression of $C/T$ is about 39%, significantly less than the essentially complete suppression of the thermopower $Q$. $Q$ is suppressed to a different degree when the field is applied in different directions with respect to the temperature gradient along the $CoO_2$ layers. In any case, $C(T)$ is much less suppressed than $Q(T)$. This is somewhat surprising as $Q$ is given by the entropy transported by the charge carriers. Apparently more than one type of carriers has to be considered. It is instructive to make a comparison with the magnetoresistance measured on a crystal with the current parallel to the $CoO_2$ layers and the magnetic field perpendicular to the layers. (The same geometry as for the upper $\Delta Q(T)$ curve). Of foremost interest is the fact that the resistance changes only very little, about an order of magnitude less than the other properties. Thus the zero-field rapid sublinear increase with $T$ remains essentially unaffected by a field. While $R$ increases at 10 K by 0.8%, $Q$ changes by ~11%, and $C$ by ~6%. As observed for other physical properties, there appear two temperature regions with different magnetoresistance characteristics.

Taken together, the various low temperature transport and thermal properties present a rather complex and novel situation. The key findings are: (1) the zero-field resistance with the marked sub-linear $T$ dependence indicates a low residual resistance, but a rapidly growing scattering rate as $T$ increases. $R(T)$ is only weakly affected by a magnetic field. (2) The specific heat is suppressed in a 14 T field by up to ~ 39%, which is comparable to the zero-field upturn in $C/T$ below ~ 5 K. (3) The entropy transported per unit charge measured in the thermopower $Q$ is essentially fully suppressed by a magnetic field.

The various results indicate that the low energy physics of $Na_xCoO_2$ (at least near $x$ ~0.7) and the various excitations probed in the experiments cannot be captured in a single band picture. Rather, one has to consider two distinct electronic components. It would appear natural to associate the "normal" component with the large hole-like band, and the "anomalous" one with the six small hole pockets, or a similar situation with two distinct bands crossing at the Fermi level [34]. While the width of the main band is strongly reduced due to correlations, and its Fermi energy is estimated to be a few tens of meV, it dominates the charge transport and it is only weakly affected by a magnetic field. The "anomalous" component is characterized by a small energy scale and is strongly affected by a magnetic field. Thus it would be of interest to study the spin and charge dynamics resulting from the interactions between the large and small parts of the Fermi surface, and to pursue the present measurements to lower temperatures in order to further elucidate the physics of strongly correlated electrons on a the triangular lattice.

Acknowledgements
We would like to thank T.M. Rice and J.L. Gavilano for helpful discussions. This study was partly supported by the Swiss National Science Foundation.

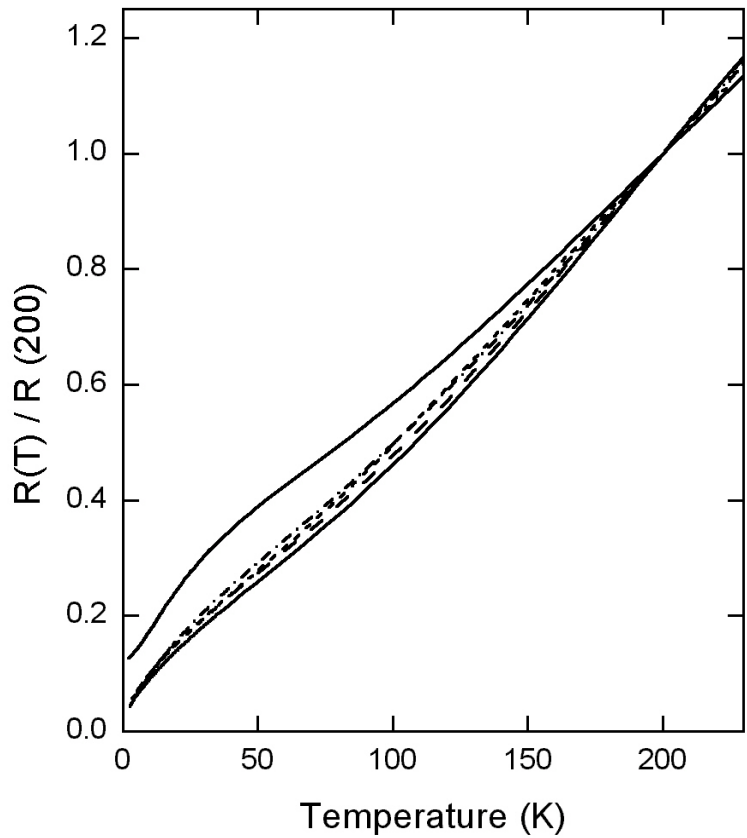

Figure 1:

Resistance of $Na_xCoO_2$ with $x=0.7$-$0.75$. The topmost curve is for a crystal with current flowing in the $CoO_2$ layer, all others for polycrystalline samples. The broad hump below ~ 100 K and the steep down-turn at lowest temperatures are noteworthy. The residual resistance is much larger in the crystal and masks the continuous drop measured for polycrystalline samples.

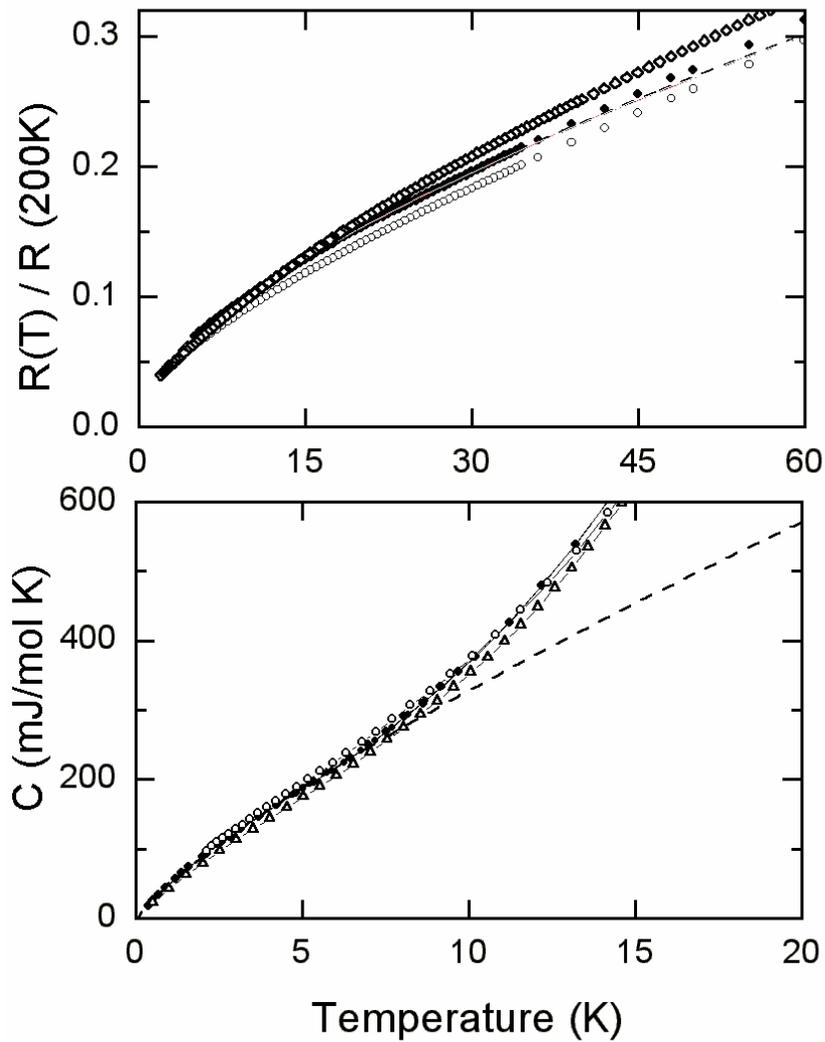

Figure 2

Sub-linear temperature dependence of the zero-field resistance and the zero-field specific heat for various samples. A power law fit $\sim T^n$ to one data set each is shown by dashed lines, with $n \sim 0.65$-$0.7$ for the resistance curves, and $n \sim 0.85$ for the specific heat. The different $T$ scales reflect the different energy windows probed when measuring the two different physical properties.

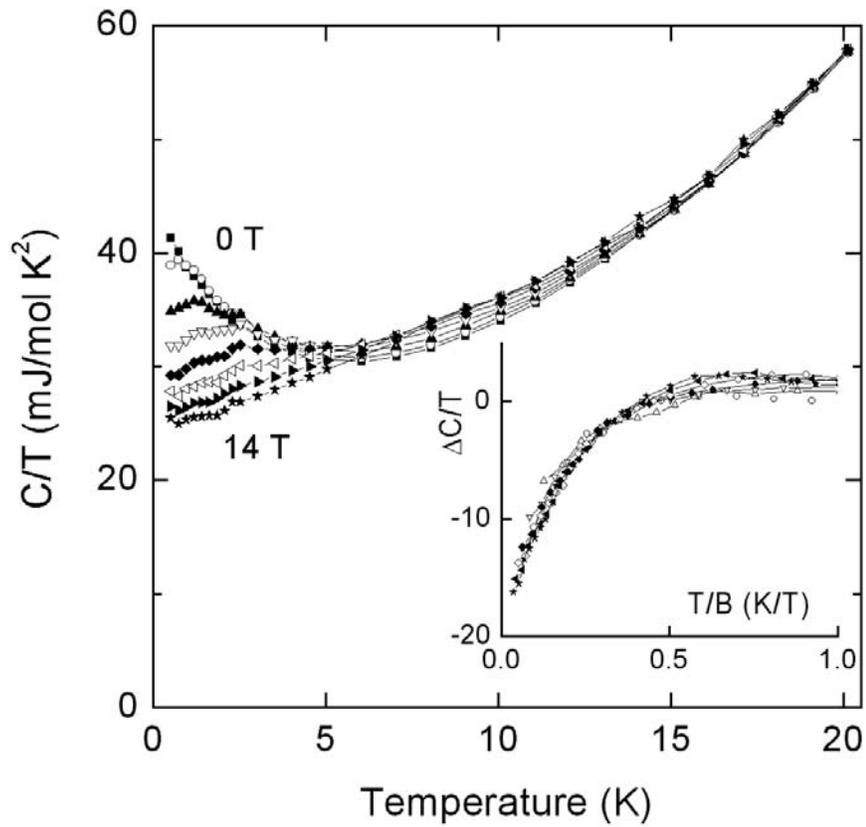

Figure 3

The low temperature specific heat in magnetic fields up to 14 T, in steps of 2 T. The suppression $\Delta C / T$ is shown in the inset on a $T / B$ scale. The curves shown are the measured data minus a low-temperature Schottky-type contribution (c.f. main text).

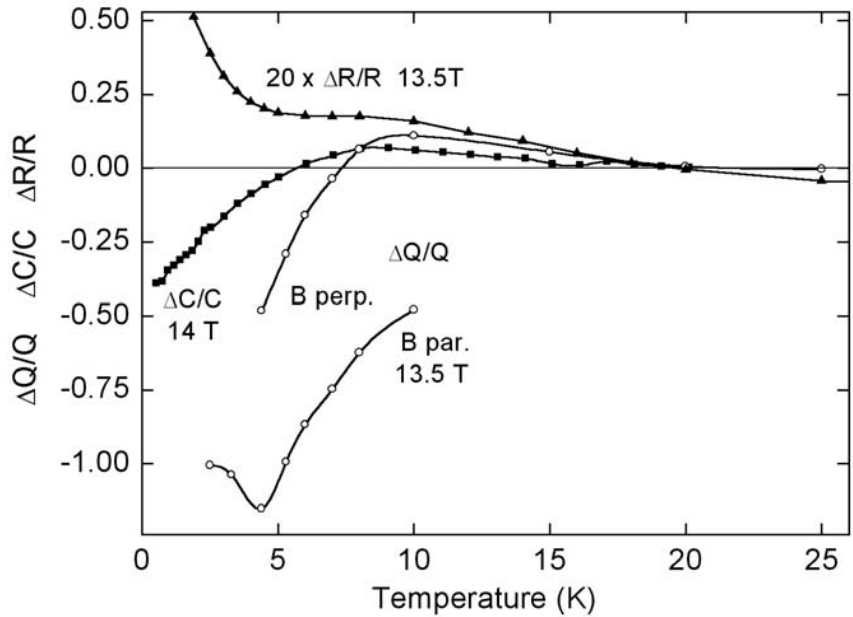

Figure 4

Influence of a magnetic field on the specific heat $C$, the thermoelectric power $Q$ and on the resistance $R$. The changes $\Delta C/C$, $\Delta Q/Q$ and $\Delta R/R$ are shown for 13.5 or 14 T, respectively. $\Delta Q/Q$ was measured on single crystals with the magnetic field parallel and perpendicular to the current flowing along the $CoO_2$ layers [22], $\Delta R/R$ in the perpendicular configuration, and $\Delta C/C$ was measured on polycrystalline samples.